\documentclass{vgtc}                          

\ifpdf
  \pdfoutput=1\relax                   
  \usepackage{graphicx}                
  \DeclareGraphicsExtensions{.pdf,.png,.jpg,.jpeg} 
\else
  \usepackage{graphicx}                
  \DeclareGraphicsExtensions{.eps}     
\fi%

\graphicspath{{figures/}{pictures/}{images/}{./}} 

\usepackage{microtype}                 
\PassOptionsToPackage{warn}{textcomp}  
\usepackage{textcomp}                  
\usepackage{mathptmx}                  
\usepackage{times}                     
\usepackage{svg}
\usepackage{cite}                      
\usepackage{tabu}                      
\usepackage{booktabs}                  

\onlineid{0}

\vgtccategory{Research}

\vgtcinsertpkg

\title{An Image-Space Split-Rendering Approach to Accelerate Low-Powered Virtual Reality}

\author{Ville Cantory\thanks{e-mail: canto063@umn.edu}\\ %
        \scriptsize University of Minnesota %
\and Nathan Ringo\thanks{e-mail: ringo025@umn.edu}\\ %
     \scriptsize University of MInnesota. }

\teaser{
  \centering
  \includegraphics[scale = 0.15]{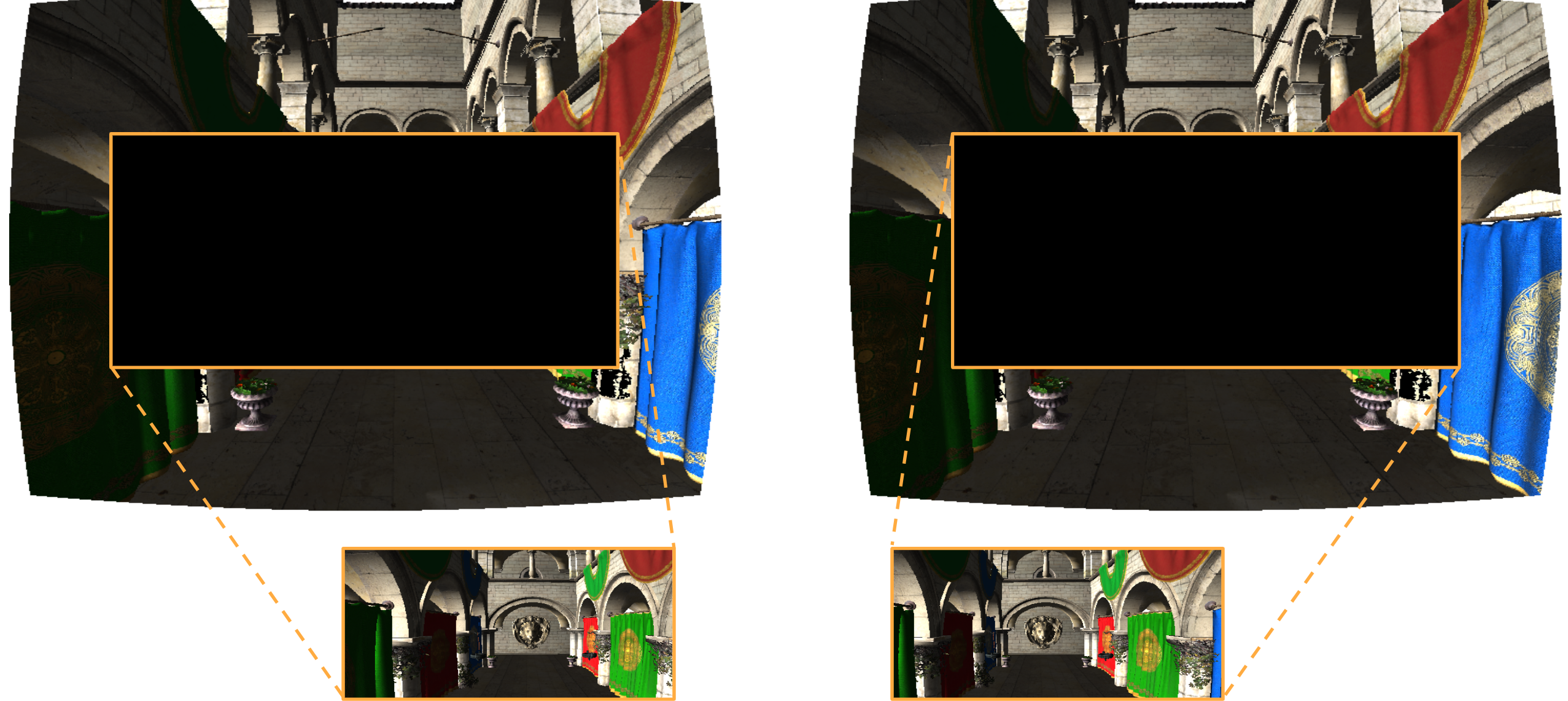}
  \caption{Illustration of the proposed client-server workload sharing model. In this example, the client (top) renders only the peripheral pixels into a reduced resolution buffer, while the server (bottom) renders only the center pixels, which are streamed to the client and merged on the client in real-time fashion.}
  \label{fig:teaser}
}

\abstract{
Current mobile GPUs lack the computing power needed to support high end VR rendering. 
As an alternative, frames can be rendered on a server and streamed to a thin client, but this approach can incur high end-to-end latency due to processing and network requirements. 
We propose a networked split-rendering approach using an image-space division of labour between a server and a heavy client to achieve faster end-to-end image presentation rates on the mobile device while preserving image quality.
} 

\CCScatlist{
  \CCScatTwelve{Computing methodologies}{Computer graphics}{Graphics systems and interfaces}{Virtual reality};
}

\begin{document}


\firstsection{Introduction}

\maketitle

High performance virtual reality (VR) experiences requires a wired connection from the head mounted display (HMD) system to a high powered desktop.
In contrast, standalone systems containing a mobile-class System on a Chip (SoC) can provide wireless VR experiences, but they do not have the compute capabilities to render the highest quality graphics for VR applications.

Some methods split the rendering pipeline across the server and a thin client \cite{mueller2018shading, hladky2021snakebinning, li2021log}, with the thin client doing much less of the processing than the server.
A heavy client, however, will incur lower communication costs with the server, decreasing encoding and network overhead.
Screen partitioning methods for parallel rendering have been explored in multi-GPU multi-display workstations to balance the number of vertices rendered by each GPU \cite{dong2019screen}, but these partitions rely on object-space information.

We propose a split-rendering solution that uses an image-space partitioning of the scene coupled with a client-heavy approach to avoid high overhead costs.
Our system uses a locally networked server to share the workload with a mobile client in rendering the full scene.
The client and server each execute the full rendering pipeline to draw separate portions of the framebuffer.
The server then uses lossless compression to encode its rendered subregion and sends it to the client, which decodes the subframe and merges the two together.
In this paper, we introduce the proposed split rendering method, showcased in a functioning system, as well as an evaluation on end-to-end latency.

\section{System}
Our system uses a heavy client to mitigate high overheads from server-side processing and network latency that can be incurred by thin-client systems.
The client's workload is reduced by rendering a subset of a frame, with the server rendering the remaining portion.
The full server-client system architecture is shown in Figure 2.

To avoid artifacting, lossless compression is used. 
As this increases the encoding time compared to lossy compression, it is advantageous for the server to render a small region at a high sampling rate, while the client can render the remaining pixels at a reduced sampling rate, akin to Fixed Foveated Rendering (FFR).
We opt to have the server render the center (ie, foveal) pixels while the client renders the peripheral pixels.
The viewport is synced across both devices, with the framebuffer being partitioned across them; additionally, the client and server are kept in lockstep, with the server idling until the client displays the final frame and sends updated camera pose information to the server.

\setlength{\belowcaptionskip}{-10pt}
\begin{figure}
    \centering
    \begin{minipage}[t]{\columnwidth}
        \includegraphics[width = \linewidth]{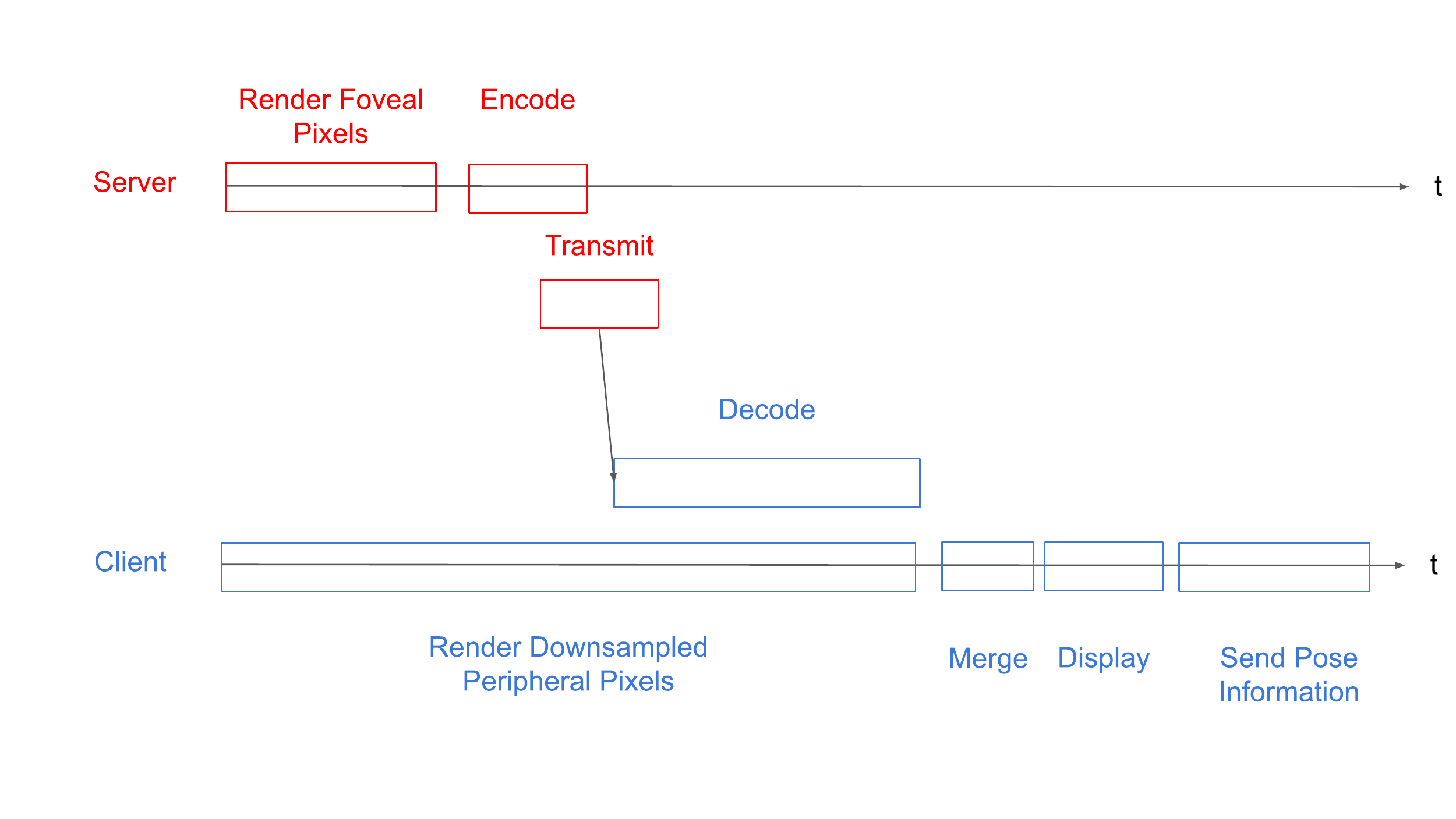}
        \caption{Architecture and ideal timeline for split-rendering. The server renders the center pixels, encodes them into an H.264 video frame, and sends the video frame to the client. The client receives that video frame asynchronously with its own rendering of the remaining pixels, decodes the video frame, and merges the two together to be displayed in one final image for the user. Often, the decoding stage finishes after the client's rendering, delaying the merge and display stages}
    \end{minipage}
    \label{fig:architecture}
\end{figure}

At the start of the frame, the server and client have the same camera pose.
The server renders its foveal pixels at the full sampling rate, and then undergoes a lossless H.264 encoding stage for each eye.
Our implementation uses CPU encoding / decoding.
Once each eye has been encoded, it is transmitted over a TCP connection to the client.
Meanwhile, the client renders its peripheral pixels at a reduced sampling rate, while another thread receives and decodes the encoded video frame from the server.   
Once the client's rendering and decoding has finished, the server's subframe can be merged with the client's subframe and displayed to the user. Updated pose information is then sent to the server, completing an entire frame.
Multiview rendering is implemented with the VK\_KHR\_multiview Vulkan extension.

\section{Evaluation}
To evaluate our approach, we compute the median end-to-end latency (time to complete one frame), showcased in Figure 3.
Additional benchmarks for each stage of the server and client's architecture are also shown in Tables 1 and 2.
The full resolution of each frame is 2400x1080 pixels, giving a resolution of 1200x1080 per eye. 
Each foveal region rendered by the server is 512x360 pixels per eye.
A moving camera pass through Crytek's Sponza (262k triangles, 49 textures) is evaluated over 1000 frames.
The client is a Oneplus 8 with a Snapdragon 865 SoC with an Adreno 650 GPU.
The server has an Intel i7-6850K CPU and an NVIDIA GTX 1080 GPU, and is connected via a 1Gbit/s ethernet cable to a Netgear Nighthawk R8500.
The client is connected to the router over 5Ghz WiFi.

The center pixels are rendered at the full sampling rate, and the peripheral pixels are sampled into a 1440x648 reduced resolution buffer.
In the split-rendering case, the client only renders the peripheral pixels with the reduced resolution buffer, while the server renders the center pixels at the full sampling rate and sends them to the client.
Using this method, the native phone client rendering the full scene on its own rendered at a median of 32.2 ms/frame (31 fps, \textit{IQR} = 7.797), and the split rendering solution rendered at 26.17 ms/frame (38 fps, \textit{IQR} = 4.572), improving the end-to-end latency per frame by 23.05\%.

\setlength{\belowcaptionskip}{-10pt}
\begin{figure}
    \centering
    \begin{minipage}[t]{\columnwidth}
        \includegraphics[width = \linewidth]{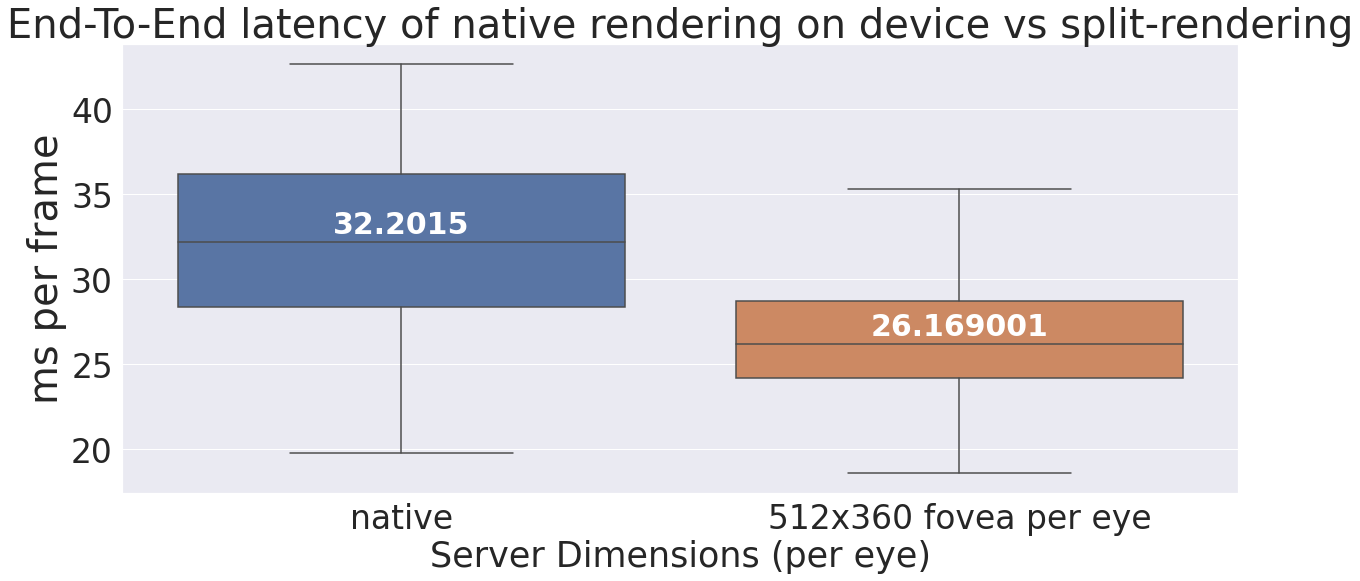}
        \caption{Median end-to-end runtimes in ms/frame comparing the runtime for native rendering (phone only) against split-rendering where the server renders a 512x360 subframe per eye and streams to the client, with the client rendering the remaining pixels into a reduced resolution buffer. There is a 23.05\% decrease in end-to-end latency using the split-rendering method.}
    \end{minipage}
    \label{fig:architecture}
\end{figure}

\begin{table}
\centering
  \caption{Client profiling on Sponza for our split-rendering model where the server renders the foveal regions per eye at full resolution, and the client renders the peripheral pixels at a reduced resolution. Updating the camera pose typically took $<$ 0.6 ms. All times are median values, reported as milliseconds except for Mbps (Megabits per second).}
  \label{tab:freq}
  \begin{tabular}{ccccccl}
    \toprule
    Server Dims&Network&Decode&Merge&Mbps \\
    \midrule
    512x360 & 4.99 & 18.64 & 0.96 & 707.83 \\
  \bottomrule
\end{tabular}
\end{table}

\begin{table}
\centering
  \caption{Server profiling when rendering for the OnePlus 8 phone client. The full resolution for the phone is 2400x1080,  with the server rendering 1024x360 pixels (512x360 per eye) at the full sampling rate.}
  \label{tab:freq}
  \begin{tabular}{ccccl}
    \toprule
    Server Dims&Draw Time&Encode Time \\
    \midrule
    512x360 & 4.42 & 15.33 \\
  \bottomrule
\end{tabular}
\end{table}

\section{Conclusion \& Future Work}
Our experiments have shown that an image-space subdivision of labour wherein a locally networked server renders a portion of the center pixels while the mobile client renders the remaining peripheral pixels at a reduced resolution can lead to large performance improvements over native rendering on the mobile client exclusively.
Our results showed a speedup of 23.05\% using our image-space split-rendering method.

This work can be extended in several ways: The workloads could be swapped, with the server employing lossy compression to render the peripheral regions while the client renders the fovea, at the cost of visual artifacts appearing in the periphery.
Mueller et. al showed that reshading a particular pixel often does not need to occur between frames \cite{mueller2021temporally}.
DeepFovea showed that with adequate hardware, a small portion of pixels can be sampled to fully reconstruct a scene \cite{kaplanyan2019deepfovea}, which could allow the client to be less reliant on receiving fully rendered subframes from the server.

\bibliographystyle{abbrv-doi}

\bibliography{main}
\end{document}